\documentclass[twocolumn,showpacs,preprintnumbers,amsmath,amssymb,prb]{revtex4}

\usepackage{graphicx}
\usepackage{dcolumn}
\usepackage{bm}

\begin{document}

\input epsf.sty

\draft
\widetext

\title
{
Magnetic and superconducting phase diagram of electron-doped Pr$_{1-x}$LaCe$_{x}$CuO$_{4}$
}









\author{M. Fujita$^{1}$}
\email{fujita@scl.kyoto-u.ac.jp}
\author{T. Kubo$^{1}$}%
\author{S. Kuroshima$^{1}$}%
\author{T. Uefuji$^{1}$}%
\author{K. Kawashima$^{1}$}%
\author{K. Yamada$^{1}$}%
\author{I. Watanabe$^{2}$}%
\author{K. Nagamine$^{2,3}$}%

\affiliation{%
$^{1}$Institute for Chemical Research, Kyoto University, Uji, Kyoto 610-0011, Japan
}

\affiliation{%
$^{2}$RIKEN (The Institute of Physical and Chemical Research), Wako, Saitama 351-0198, Japan
}

\affiliation{%
$^{3}$Meson Science Laboratory, Institute of Materials Structure Science, High Energy Accelerator Research Organization (KEK-MSL), Tsukuba, Ibaraki 305-0801, Japan
}

\date{\today}




\begin{abstract}
We have investigated the magnetism and the superconductivity of the electron-doped Pr$_{1-x}$LaCe$_{x}$CuO$_{4}$ (PLCCO) by means of zero-field muon spin rotation/relaxation (ZF-$\mu$SR) and magnetic susceptibility measurements. 
At low temperatures, a well-defined muon spin rotation free from the effect of rare earth moments was observed for samples with {\it x}$\leq$0.08 corresponding to the antiferromagnetic (AF) order of Cu spins. 
%
%
%
Bulk superconductivity was identified in a wide Ce concentration range of 0.09$\leq${\it x}$\leq$0.20 with a maximum transition temperature of 26 K. 
Abrupt appearance of the superconducting (SC) phase at {\it x}$\sim$0.09 is concomitant with a destroy of the AF ordered phase, indicating the competitive relation between two phases. 
Possible relation between the wide SC phase and the lattice spacing is discussed. 
\end{abstract}



\maketitle

%
Electron-hole symmetry of a pairing mechanism is one of the central issues in a research on high-{\it T}$_{c}$ superconductivity. 
It is widely believed that a universal role of magnetism exists because either type of carrier doping into Mott insulators induces superconductivity. 
For understanding the relationship between magnetism and superconductivity, electronic phase diagrams provide important clues. 
In the hole-doped La$_{2-x}$Sr$_x$CuO$_4$ (LSCO) system, antiferromagnetic (AF) and superconducting (SC) phases are well separated: SC phase exists in a wide range of 0.06$\leq${\it x}$\leq$0.27 with a parabolic doping dependence of the SC transition temperature, {\it T}$_{c}$, while AF phase is located in a narrow range of {\it x}$\leq$0.02.~\cite{Takagi89,Torrance} 
In contrast, in the electron-doped Nd$_{2-x}$Ce$_x$CuO$_4$ (NCCO) and Pr$_{2-x}$Ce$_x$CuO$_4$ (PCCO) systems, the optimum superconductivity adjoins a broad AF phase (0$\leq${\it x}$\leq$0.14) and the SC phase exists in a narrow doping range of 0.14$\leq${\it x}$\leq$ 0.18.~\cite{Takagi,Luke_nature,Luke} 
Therefore, it is important to clarify the origin of electron-hole doping asymmetry seen in the phase diagram and to reveal the universal feature in the relation between  magnetism and  superconductivity irrespective of types of carrier. 

On the other hand, recent intensive studies on the LSCO system by $\mu$SR~\cite{Weidinger,Niedermayer}, nuclear magnetic resonance~\cite{Julien} and neutron scattering~\cite{Fujita} techniques have claimed that a short-range AF ordered phase above {\it x}=0.02 persists in under-doped regions, and therefore coexists or phase separates with superconductivity. 
This penetration seems to be a contrastive feature with an incompatible relation between AF and SC phases in the electron-doped systems, suggested in the aforementioned phase diagram. 
Universal feature in the phase diagram of electron-doped system, however, is still controversial due to the limited number of comprehensive studies on both magnetism and superconductivity, and the difficulties in preparing samples, especially single crystals.\cite{Kurahashi}

In this paper, we present the phase diagram of the electron-doped PLCCO system\cite{Isawa} over a wide Ce concentration range obtained by ZF-$\mu$SR and magnetic susceptibility measurements. 
An advantage of this system is that the SC phase is extending to a lower doping region compared to the case of PCCO system.~\cite{Koike,Garcia} 
Thus, by investigating the magnetic phase in the system, the relation between AF and SC phases can be clarified. 
Furthermore, compared to the NCCO system, a considerably smaller effect of the rare-earth moment is suitable for studying the inherent nature of Cu$^{2+}$ spins. 
Present study yields important information on the relation between the AF and SC phases: 
(i) Upon Ce doping the AF ordered phase is drastically suppressed at {\it x} $\sim$0.09 where the SC phase abruptly appears at the ground state, suggesting a  competition between the two phases, and 
(ii) The SC region of the PLCCO (0.09$\leq${\it x}$\leq$0.20) is extending to {\it both lower and higher doping region} compared with that of PCCO (0.14$\leq${\it x}$\leq$0.18). 
%
%

Single crystals ({\it x}=0.08, 0.09, 0.11, 0.13, 0.15, 0.17, 0.18 and 0.20) and powder samples ({\it x}=0.04, 0.06, 0.09 and 0.11) are grown using a traveling-solvent floating-zone method and a solid-state reaction, respectively. 
All samples are carefully annealed under argon gas flow at 900-950 $^{\circ}$C for $\sim$10 h and single crystals are subsequently annealed under O$_{2}$ gas-flow at 500 $^{\circ}$C for $\sim$10 h. 
Removed oxygen content per unit formula from as-grown samples was determined to be 0.03-0.05 from the weight loss of the sample after the annealing treatment. 
%
%
%
For the characterization of samples, we examined the Ce content and the lattice constants by the inductively coupled plasma (ICP) spectrometer and the X-ray powder diffractometer, respectively. 
Evaluated Ce concentrations are approximately the same as the nominal concentrations. 
\linebreak
\begin{figure}[t]
\centerline{\epsfxsize=2.6in\epsfbox{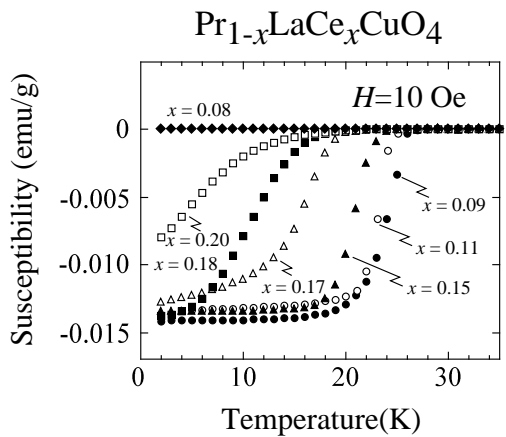}}
\caption
{
Magnetic susceptibility measured for single crystal samples of the PLCCO system after the zero-field-cooling process. 
}
\end{figure}
%
\noindent
At room temperature both {\it a} and {\it c}-axis lattice constants, which are larger than those in PCCO\cite{Kuroshima}, change monotonically with {\it x}. 
Details of the sample preparation and the characterization will be presented in a separate paper.\cite{Fujita_unpub}

In order to determine {\it T$_c$}, we measured the diamagnetic susceptibility by a SQUID magnetometer. 
Figure 1 shows the susceptibility for the annealed single crystal samples in an applied field of 10 Oe after the zero-field-cooling process. 
SC transitions are observed in the wide Ce concentration range of 0.09$\leq${\it x}$\leq$0.20, while no bulk superconductivity is detected for 0.08$\leq${\it x} samples. 
Based on these results the lower critical concentration of the bulk superconductivity is estimated to be between {\it x}=0.08 and 0.09. 
Note that onset {\it T}$_{c}$$^{\prime}$s of the {\it x}=0.09 and 0.11 powder samples are identical with those of single crystals and superconducting transition was not observed in the 0.04 and 0.06 samples. 
Thus, we concluded that the phase diagram for {\it x}$\leq$0.11 can be well characterized by using either powder or single crystal samples. 
We also note that the uniform magnetic susceptibility of PLCCO above {\it T}$_{c}$ is only $\sim$$1\%$ of that of NCCO, demonstrating an advantage of the PLCCO system for elucidation of magnetic properties of Cu spin. 

$\mu$SR measurements are performed on powder ({\it x}=0.04, 0.06, 0.09 and 0.11) and single crystal ({\it x}=0.08) samples at the pulsed muon source, RIKEN-RAL Muon Facility, Rutherford Appleton Laboratory in UK. 
These chosen values for {\it x} span the boundary between AF and SC phases. 
Positive surface-muons with perfectly polarized spins parallel to the beam and with the momentum of 29.8 MeV/c are implanted into sample.  
Then muon spins are depolarized by precessing around a local magnetic field at the muon sites. 
Therefore, the time evolution of muon spin polarization ($\mu$SR time spectrum) obtained by the asymmetry of the decay positron emission rate between forward and backward counters, {\it A}({\it t}), provides information on the distribution and/or the fluctuation of the local magnetic field and the volume fraction of the magnetically ordered phase.\cite{Hayano} 

In Fig. 2, the normalized $\mu$SR time spectra after subtracting time-independent background are shown for non-SC ({\it x}=0.08) and SC ({\it x}=0.11) samples. 
In both samples, a Gaussian depolarization is formed in the time spectra at high temperatures consistent with a static nuclear-dipole field and the rapid fluctuation of Cu$^{2+}$ spins. 
\linebreak
\begin{figure}[t]
\centerline{\epsfxsize=2.5in\epsfbox{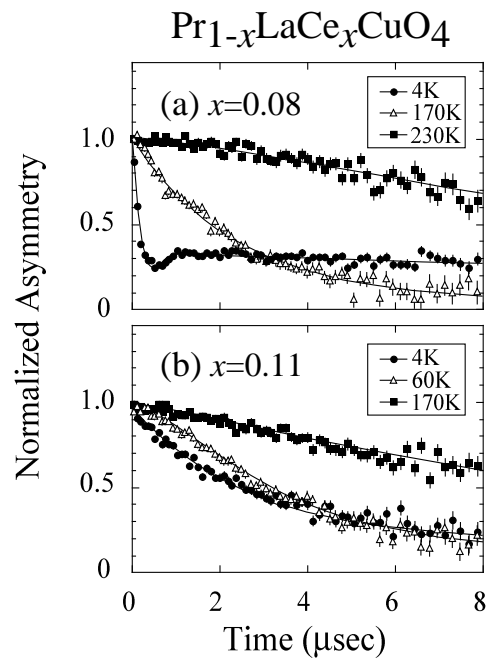}}
\caption
{
ZF-$\mu$SR time spectra of PLCCO with (a) non-superconducting ({\it x} = 0.08) and (b) superconducting ({\it x}= 0.11) samples. Solid lines are results fitted with Eqs. (1) and (2). (See text.)
}
\end{figure}
\noindent
At lower temperatures, the time spectra change from a Gaussian-type depolarization to an exponential one in both samples. 
This change suggests the appearance of static or quasistatic internal magnetic field at muon sites possibly due to the development of Cu spin correlation and the slowing down of spin fluctuations. 
Upon further cooling to 4K, an additional muon spin rotation corresponding to the magnetic order appears in the {\it x}=0.08 sample, while such a clear rotation is not observed in the {\it x}=0.11 sample. 
Therefore, magnetic property changes near the phase boundary upon electron-doping. 
Flat time spectrum beyond $\sim$1 $\mu$sec is clearly seen in {\it x}=0.08 sample at 4K, meaning a negligible effect of the Pr spin fluctuation in contrast to the case of NCCO.\cite{Uefuji,Watanabe} 
%
%
%

For the qualitative analysis of the time spectra, we first assumed a combination of Gaussian and exponential functions:
%
\begin{equation}
   {\it A}(\it t)={\it A}_{1}{\it exp}(-{\lambda}_{1}t)+{\it A}_{2}{\it exp}(-{\lambda}_{2}^{2}t^{2}),
\hspace{5mm}
\end{equation}
%
with {\it A}$_{1}$+{\it A}$_{2}$=1, where {\it A}$_{1}$ and $\lambda$$_{1}$, {\it A}$_{2}$ and $\lambda$$_{2}$ are the initial asymmetry at {\it t}=0 and the depolarization rate for the exponential and Gaussian components, respectively. 
\linebreak
\begin{figure}[t]
\centerline{\epsfxsize=2.3in\epsfbox{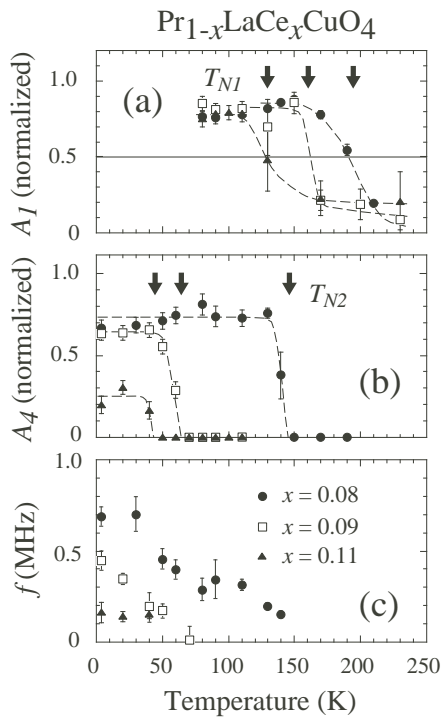}}
\caption
{
Fitting parameters of (a) initial asymmetry {\it A}$_{1}$ of the exponential component in Eq. (1), (b) {\it A}$_{2}$ and (c) the frequency {\it f} of the rotation component in Eq. (2) for {\it x} = 0.08, 0.09 and 0.11 samples. Dashed lines are guides to the eye. 
}
\end{figure}
\noindent
The time spectra at the higher temperatures ($\geq$80 K) are well reproduced by this function. 
(In Figs. 2(a) and (b), solid lines for the time spectra at the highest temperature are the fitted results by Eq. (1). )
With decreasing the temperature, {\it A}$_{1}$ increases due to the development of spin correlation. 
We defined a characteristic temperature as {\it T}$_{N1}$ where {\it A}$_{1}$ exceeds 0.5 or the exponential component dominates. (See Fig. 3(a).) 
%
Then, to get more information regarding the ordered phase at low temperatures, the time spectra below {\it T}$_{N1}$ were again fitted to the following equation:
%
\begin{equation}
   {\it A}(\it t)={\it A}_{3}{\it exp}(-{\lambda}_{3}t)+{\it A}_{4}{\it exp}(-{\lambda}_{4}t)cos(2\pi{\it ft}+\phi),
\hspace{5mm}
\end{equation}
%
%
%
with {\it A}$_{3}$+{\it A}$_{4}$=1, where the first and second terms express components of relaxation and rotation of muon spin. 
The parameters {\it A}$_{3}$ and $\lambda$$_{3}$ are the initial asymmetry and the depolarization rate of exponential relaxation, respectively. 
{\it A}$_{4}$ and $\lambda$$_{4}$ are those of rotation component and {\it f} and $\phi$ are the frequency and the initial phase of rotation. 
Solid lines for the time spectra at lower two temperatures in Figs. 2(a) and (b) are the fitted results by Eq. (2). 

In Figs. 3(b) and (c), the obtained parameters of {\it A}$_{4}$ and {\it f}, which represent the AF volume fraction and the relative internal magnetic field at the muon site, respectively, are shown for the samples located near the boundary. 
We define {\it T}$_{N2}$ as the onset temperature for the appearance of muon spin rotation, corresponding to the existence of {\it static} AF ordered state. 
At low temperature, both {\it A}$_{4}$ and {\it f} decrease as {\it x} increases. 
However, as seen in Fig. 3(c) the internal magnetic field at muon sites decreases upon electron doping possibly due to the change in either the amplitude or the direction of staggered moment of Cu spins. 
In contrast, in the hole-doped LSCO system, {\it f} in the long-range AF ordered phase is constant for {\it x}$\leq$0.02\cite{Harshman,Borsa} where the evidence of a phase separation between three-dimensional long-range-ordered phase and spin-glass phase was observed.\cite{Chou,Matsuda} 
Therefore, the AF order degrades  rather homogeneously in space upon electron-doping in contrast to the inhomogeneous degradation in the hole-doped system: In the electron-doped system, the magnetic structure and/or staggared moment are modified by doping, while in the hole-doped system, those of the undoped system persists in the slightly doped compound.\cite{Lavrov,Matsuda}  
It should be noted that magnetic Bragg peaks were observed by elastic neutron-scattering measurements below {\it T}$_{N1}$\cite{Fujita_unpub_2}, while no clear evidence for a static internal field at the temperature between {\it T}$_{N1}$ and {\it T}$_{N2}$ was obtained by longitudinal field $\mu$SR measurement.\cite{Kubo} 
Thus, individual Cu$^{2+}$ spins are fluctuating faster compared to the time scale of $\mu$SR measurement (typically 10$^{-6}$-10$^{-11}$ sec), although there exists {\it time-averaged} ordered moment. (The state with time-averaged moment is defined to be N$\acute{e}$el state. ) 
\linebreak
\begin{figure}[t]
\centerline{\epsfxsize=2.77in\epsfbox{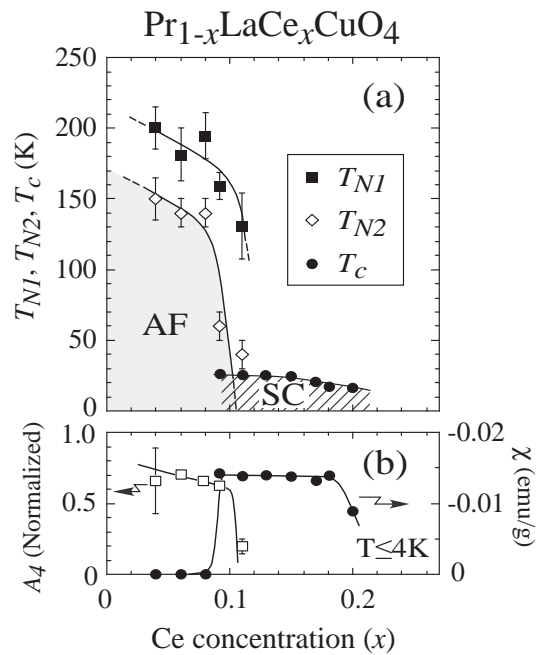}}
\caption
{
Doping dependence of (a) {\it T}$_{N1}$ (closed squares), {\it T}$_{N2}$ (open diamonds) and {\it T}$_{c}$(onset) (closed circles) and (b) initial asymmetry for the rotation component of {\it A}$_{4}$ (open squares) and diamagnetic susceptibility (closed circles) at temperature below 4 K. Solid lines are guides to the eye. Shaded and hatched areas in the upper figure correspond to AF and SC phases, respectively. 
}
\end{figure}
\noindent

In Fig. 4 (a), the doping dependence of {\it T}$_{N1}$, {\it T}$_{N2}$ and {\it T}$_{c}$(onset) are summarized. 
Upon electron doping, bulk superconductivity with optimum {\it T}$_{c}$ of 26 K abruptly appears at {\it x}$\sim$0.09 like a first order-transition and {\it T}$_{N2}$ is dramatically suppressed at the same time. 
Therefore, at the ground state, SC phase appears concomitant with the disappearance of the AF ordered phase as seen in the NCCO system,\cite{Luke,Uefuji} although two phases are partially overlapped due to coexistence or microscopic phase separation. 
This result combined with a relation between the doping dependences of {\it A}$_{4}$ and the diamagnetic susceptibility, $\chi$, at low temperatures (Fig. 4(b)) clearly demonstrates a competitive relation between AF and SC phases. 
Abrupt onset of optimum superconductivity accompanied by the disappearance of AF phase is different from the result in the hole-doped system showing a penetration of short-range AF ordered phase into the under-doped SC one.~\cite{Weidinger,Niedermayer,Julien,Fujita} 
%
%

Now we turn to the discussion on the doping range of SC phase. 
As seen in Figs. 4(a), {\it T}$_{c}$ is insensitive to the Ce concentration and this feature characterizes the wide SC phase. 
The wide SC phase would be related with the increase of effective carriers by La substitution suggested from resistivity and Seebeck coefficient measurements.\cite{Koike,Garcia} 
Arima {\it et al}. reported a reduction of charge-transfer (CT) energy between Cu 3{\it d} and O 2{\it p} bands as stretching Cu-O bond, i.e. lattice spacing.\cite{Arima} 
If the reduction of the CT energy is greater than the loss from decreasing orbital overlap\cite{Manthiram}, then the gain in mobility would make the introduction of electrons into the CuO$_{2}$ plane easier, resulting in the wider SC phase. 
In other words, the narrower SC phases in NCCO and PCCO compared to that in PLCCO originate from the short lattice spacing. 

On the other hand, for the appearance of superconductivity in the 2-1-4 electron-doped systems, a reduction procedure such as heat treatment is necessary. 
%
%
Brinkmann {\it et al}. reported an extension of SC phase in PCCO by an improved reduction technique and Kurahashi {\it et al}. shows an occurrence of optimum superconductivity in the Nd$_{1.85}$Ce$_{0.15}$CuO$_{4}$ by an adequate heat treatment.\cite{Brinkmann,Kurahashi} 
Therefore, SC composition range depends on the reduction procedure. 
The slight difference in the SC composition range for present crystals and powder samples of PLCCO system\cite{Garcia,Isawa} possibly relates with differences in the heat treatment. 
%
%
%
%
In order to clarify the fundamental features in the electron doped superconductivity, further comprehensive studies on single crystals are required. 

In conclusion, we have performed $\mu$SR and magnetic susceptibility measurements for the electron-doped PLCCO system. 
AF order was observed in the sample with 0.04$\leq${\it x}$\leq$0.11. The AF order was dramatically suppressed at {\it x} $\sim$0.09 which corresponds to the onset of the significantly wide SC phase (0.09$\leq${\it x}$\leq$0.20) upon doping. 
The obtained phase diagram combined with the doping dependences of AF volume fraction and internal magnetic field clearly demonstrates a competitive relation between AF and SC phases. 

We thank M. Kofu and K. Hirota for technical assistance of ICP measurements at Tohoku University, and Isawa, G-q. Zheng and M. Matsuda for helpful discussion. 
This work was supported in part by the Japanese Ministry of Education, Culture, Sports, Science and Technology, Grant-in-Aid for Scientific Research on Priority Areas (Novel Quantum Phenomena in Transition Metal Oxides), for Scientific Research (A), for Encouragement of Young Scientists, and for Creative Scientific Research "Collaboratory on Electron Correlations - Toward a New Research Network between Physics and Chemistry -", by the Japan Science and Technology Corporation, the Core Research for Evolutional Science and Technology Project (CREST).



\end{document}